\documentclass[prd,showkeys,floatfix,twocolumn,amsmath,amssymb,floatfix]{revtex4}
\usepackage{graphicx}
\usepackage{multirow}
\usepackage{subfigure}
\usepackage[colorlinks,citecolor=blue,anchorcolor=red,menucolor=red,linkcolor=red,filecolor=red,runcolor=red,urlcolor=blue,frenchlinks=red]{hyperref}
\usepackage{enumitem}
\usepackage{slashed}
\usepackage[normalem]{ulem}

\makeatletter
\renewcommand{\@thesubfigure}{\hskip\subfiglabelskip}
\makeatother
\allowdisplaybreaks[3]

\begin{document}
%=====================================================================================
\title{Improved prediction of the mass splitting for $P$-wave $\Omega$ baryons}
%=====================================================================================
\author{Niu Su$^{1,2}$}
\email{suniu@tyut.edu.cn}
\author{Hua-Xing Chen$^3$}
\email{hxchen@seu.edu.cn}
\author{Philipp Gubler$^4$}
\email{philipp.gubler1@gmail.com}
\author{Atsushi Hosaka$^{2,4}$}
\email{hosaka@rcnp.osaka-u.ac.jp}

\affiliation{$^1$College of Physics and Optoelectronic Engineering, Taiyuan University of Technology, Taiyuan 030024, China\\
$^2$Research Center for Nuclear Physics (RCNP), Osaka University, Ibaraki 567-0047, Japan\\
$^3$School of Physics, Southeast University, Nanjing 210094, China\\
$^4$Advanced Science Research Center, Japan Atomic Energy Agency (JAEA), Tokai 319-1195, Japan
}
\begin{abstract}
Using the QCD sum rule method, we investigate the mass splitting for the spin-orbit partner states of the $\Omega(2012)$ baryon assuming that it is a $P$-wave excitation with $J^P=3/2^-$. This study is an extension of the previous work~\cite{Su:2024lzy} in which the masses of these states were estimated with uncertainties too large to extract the reliable mass splitting.  In the present study, by directly formulating a sum rule for the mass splitting, we obtain an improved prediction, $\delta M = M_{3/2^-} - M_{1/2^-} = -18.0^{+ 33.6}_{-17.1}$ MeV. This result provides a more quantitative insight into the spectrum of $P$-wave $\Omega$ baryons and serves as a useful reference for future experiments.
\end{abstract}
\keywords{Mass splitting, $P$-wave $\Omega$ baryons,  QCD sum rules}

\maketitle

\section{Introduction}
\label{sec:intro}

Hadrons are composed of quarks and gluons, with their strong interactions governed by Quantum Chromodynamics (QCD). The quark model systematically classified hadrons into mesons, made of a quark-antiquark pair, and baryons, composed of three quarks~\cite{Gell-Mann:1964ewy,Zweig:1964jf}. This theoretical framework has achieved remarkable success in explaining hadron properties. Experimentally, all ground-state light-flavor mesons and baryons consisting of $u$, $d$, and $s$ quarks have been established~\cite{ParticleDataGroup:2024cfk}, and a large number of their excited states have also been observed. However, the existence and structure of some highly excited states remain controversial, potentially providing crucial insights into the non-perturbative features of QCD in the low-energy regime~\cite{Chen:2016qju,Hosaka:2016pey,Richard:2016eis,Lebed:2016hpi,Esposito:2016noz,Guo:2017jvc,Ali:2017jda,Olsen:2017bmm,Liu:2019zoy,Brambilla:2019esw,Meng:2022ozq,Chen:2022asf,Liu:2024uxn,Wang:2025sic}.

In the $\Omega$ baryon sector, the Particle Data Group (PDG) currently lists only five states. Besides the ground-state $\Omega(1672)$, $\Omega(2012)$ and $\Omega(2250)$ are rated with three stars, indicating likely existence but requring further experimental confirmation and theoretical investigation, while $\Omega(2380)$ and $\Omega(2470)$ have two stars, reflecting weaker evidence. The present moment provides a timely opportunity to study the $\Omega$ baryon spectrum. Among these, $\Omega(2012)$ has attracted considerable attention since its discovery by the Belle experiment in 2008~\cite{Belle:2018mqs,Belle:2021gtf,Belle:2022mrg,Aliev:2018syi,Aliev:2018yjo,Polyakov:2018mow,Wang:2018hmi,Xiao:2018pwe,Liu:2019wdr,Arifi:2022ntc,Menapara:2021vug,Wang:2022zja,Zhong:2022cjx,Lin:2018nqd,Valderrama:2018bmv,Pavao:2018xub,Huang:2018wth,Gutsche:2019eoh,Ikeno:2020vqv,Zeng:2020och,Lu:2020ste,Liu:2020yen,Ikeno:2022jpe,Hu:2022pae,Lu:2022puv,Xie:2024wbd,Hockley:2024aym,Luo:2025cqs,Han:2025gkp}. In the past year, both the BESIII and ALICE experiments have reported new supporting evidence~\cite{BESIII:2024eqk,ALICE:2025atb} and the BESIII Collaboration has also observed a potential new state $\Omega(2109)$. In our previous study~\cite{Su:2024lzy}, we analyzed the mass and structure of $\Omega(2012)$ using parity-projected QCD sum rules and found it is likely a $P$-wave excitation with $J^P=3/2^-$.  We further predicted the mass of its spin–orbit partner with $J^P=1/2^-$, they are:
\begin{eqnarray}
\label{eq:mass3}M_{3/2^-} &=& 2.05^{+0.09}_{-0.10}{\rm~GeV} \, ,
\\ 
\label{eq:mass1}M_{1/2^-} &=& 2.07^{+0.07}_{-0.07}{\rm~GeV} \, .
\end{eqnarray}
Although these results imply that mass splitting of the two spin states is not large, it remained poorly constrained due to the inherent uncertainties . To improve the precision of the mass splitting for $P$-wave $\Omega$ excitations and to explore whether $\Omega(2109)$ could correspond to the spin–orbit partner of $\Omega(2012)$, we extend our previous work by formulating a dedicated QCD sum rule for the splitting, thereby providing a more accurate theoretical prediction.

This paper is organized as follows: In Sec.~\ref{sec:current}, we present the interpolating currents of the $P$-wave $\Omega$ baryon used in the calculations. In Section.~\ref{sec:sumrule}, we perform a detailed description for formulating a QCD sum rule approach to evaluate the mass splittings. Finally, Sec.~\ref{sec:summary} concludes with a brief summary and discussion.

\section{$P$-wave $\Omega$ baryon currents}
\label{sec:current}

The $\Omega$ baryon consists of three identical $s$ quarks which obeys Fermi-Dirac statistics. For $P$-wave $\Omega$ baryons with orbital angular momentum $L=1$, there exist two spin-orbit partner states with quantum number $J^P=1/2^-$ and $J^P=3/2^-$. The detailed construction of the currents for these states can be found in our previous work~\cite{Su:2024lzy}, where the currents for the corresponding excited states are constructed as $\rho$-mode excitations:
\begin{eqnarray}
J &=& -2\epsilon^{abc} ~ [(D^\mu s_a^T) C \gamma_5 s_b]
\label{def:current1}~ \gamma_\mu s_c \, ,
\\ 
J_\mu &=& -2\epsilon^{abc} ~ [(D^\nu s_a^T) C \gamma_5 s_b] \label{def:current2}~ (g_{\mu\nu} - 
{1\over4}\gamma_\mu\gamma_\nu) s_c \, ,
\end{eqnarray}
Note that these currents can also expressed as $\lambda$-mode currents via Fierz transformation, which is the case for three identical strange quarks. In this study, we empoly them to couple to the physical states as:
\begin{eqnarray}
\langle 0 | J | \Omega; 1/2^- \rangle &=& f_{1/2^-} \gamma_5 u(q) \, ,
\\ 
\langle 0 | J_\mu | \Omega; 3/2^- \rangle &=& f_{3/2^-} u_\mu(q)\, .
\label{coupling1}
\end{eqnarray}
where $f_{1/2^-(3/2^-)}$ are the coupling strength of the currents to physical states, and $u$ and $u_{\mu}$ are the spinors for spin 1/2 and 3/2 states.

\section{QCD sum rule analyses}
\label{sec:sumrule}

In this section, we formulate the QCD sum rules to calculate the mass splitting between the $P$-wave $\Omega$ baryons. We firstly study their correlation functions:
\begin{eqnarray}
\Pi_{1/2}(q^2) &=& i \int d^4x e^{iqx} \langle 0 | {\bf T}[J(x) J^\dagger(0)] | 0 \rangle  \,,
\\
\Pi_{\mu\nu,3/2}(q^2) &=& i \int d^4x e^{iqx} \langle 0 | {\bf T}[J_{\mu}(x) J_\nu^\dagger(0)] | 0 \rangle \label{twoponit}
\\ \nonumber
&=&(\frac{q_\mu q_\nu}{q^2}-g_{\mu\nu})\Pi_{3/2}(q^2)+\cdots \, ,
\end{eqnarray} 
where we have isolated the Lorentz structure corresponding to the spin-$3/2$ component $\Pi(q^2)_{3/2}$. The function $\Pi_{\xi}(q^2)$ can then be represented by a dispersion relation:
\begin{eqnarray}
\Pi_{\xi}(q^2) = \int^\infty_{s_<}\frac{\rho_{\xi}(s)}{s-q^2-i\varepsilon}ds \, ,
\end{eqnarray}
where $\xi$ denotes either $1/2$ or $3/2$,  $\rho(s) \equiv {\rm Im} \Pi(s)/\pi$ is the spectral density, and $s_< = 9 m_s^2$ is the physical threshold.

At the hadron level, we insert a complete set of intermediate states into the correlation function and extract the corresponding spectral densities:   
\begin{eqnarray}
\rho^{\rm phen}_{1/2}(s)  &\equiv& \sum_m\delta(s-M^2_m) \langle 0| J | m\rangle \langle m| J^{\dagger} |0 \rangle \label{eq:rho}
\\ \nonumber 
&=&f_{1/2^-}^2 (\slashed{q}+M_{1/2^-}) \delta(s-M_{1/2^-}^2) 
\\ \nonumber
&&+ f_{1/2^+}^2 (\slashed{q}-M_{1/2^+}) \delta(s-M_{1/2^+}^2)
\\ \nonumber
&&+ \rm \,\,Continuum \, ,
\\ 
\rho^{\rm phen}_{3/2}(s)  &\equiv& \sum_n\delta(s-M^2_n) \langle 0| J_{\mu} | n\rangle \langle n| J_{\nu}^{\dagger} |0 \rangle \label{eq:rho}
\\ \nonumber 
&=&f_{3/2^-} (\slashed{q}+M_{3/2^-}) \delta(s-M_{3/2^-}^2) 
\\ \nonumber
&&+ f_{3/2^+}^2 (\slashed{q}-M_{3/2^+}) \delta(s-M_{3/2^+}^2)
\\ \nonumber
&&+ \rm \,\,Continuum \, ,
\end{eqnarray}
These can be rewritten into two parts with different structures, denoted as $\rho^{\rm phen}_{\xi,1(0)}(s)$: 
\begin{eqnarray}
\rho^{\rm phen}_{\xi}(s)&=&\rho^{\rm phen}_{\xi,1}(s)\slashed{q}+\rho^{\rm phen}_{\xi,0}(s)\label{phen1}\, ,
\end{eqnarray}
Finally, we extract the spectral densities corresponding exclusively to the positive- or negative-parity states as:
\begin{eqnarray}
\rho^{\rm phen}_{\xi,\mp}(s) = \sqrt{s} \rho^{\rm phen}_{\xi,1}(s)\pm \rho^{\rm phen}_{\xi,0}(s) .
\end{eqnarray}

At the quark-gluon level, we use the method of the operator product expansion (OPE) to calculate the correlation function and extract the corresponding spectral densities, denoted as $\rho^{\rm OPE}_{\xi,1(0)}$:
\begin{eqnarray}
&& \nonumber\rho^{\rm OPE}_{1/2,1}(s)  
\\ &=& \label{OPE1}
{7 s^3 \over 10240 \pi^4}-{45 m_s^2 s^2 \over 2048 \pi^4}
\\  \nonumber&&
+ \Big(-{7 \langle g_s^2 GG \rangle \over 12288 \pi^4}
+{9 m_s \langle \bar s s \rangle \over 128 \pi^2}\Big )s
\\ \nonumber &&
+ \Big ( - {13 m_s \langle g_s \bar s \sigma G s \rangle \over 128 \pi^2}
+ { 7 m_s^2\langle g_s^2 GG \rangle  \over 2048 \pi^4 }\Big )
\\ \nonumber &&
-\Big ({ \langle \bar s s \rangle \langle g_s \bar s \sigma G s \rangle \over 12}
-{7 m_s\langle g_s^2 GG \rangle \langle \bar s s \rangle \over 3072 \pi^2}
\\ \nonumber &&
-{7m_s^2\langle \bar s s \rangle^2\over32 } \Big )\delta(s)
+ \Big(- {17 \langle g_s \bar s \sigma G s \rangle^2 \over 384} 
\\ \nonumber &&
-{7 \langle g_s^2 GG \rangle \langle \bar s s \rangle^2 \over 192} + { m_s\langle g_s^2 GG \rangle \langle g_s \bar s \sigma G s \rangle \over 512\pi^2}
\\ \nonumber &&
-{ m_s^2 \langle \bar s s \rangle \langle g_s \bar s \sigma G s \rangle \over 384} \Big)\delta^\prime(s)\, ,
\\ 
&& \nonumber \rho^{\rm OPE}_{1/2,0}(s)  
\\ &=& \label{OPE0}
{3 m_s s^3 \over 1024 \pi^4}-{3\langle \bar s s \rangle s^2 \over 64 \pi^2}
- \Big( {7 m_s \langle g_s^2 GG \rangle \over 2048 \pi^4}
\\  \nonumber&&
-{17 m_s^2 \langle \bar s s \rangle \over 64 \pi^2}
+{7 \langle g_s \bar s \sigma G s \rangle \over 128 \pi^2} \Big )s
\\ \nonumber &&
+ \Big (  {83 m_s^2 \langle g_s \bar s \sigma G s \rangle \over 256 \pi^2}
+ { 5 \langle \bar s s \rangle \langle g_s^2 GG \rangle  \over 768 \pi^2 }\Big )
\\ \nonumber &&
-\Big ({3 m_s^2\langle g_s^2 GG \rangle \langle \bar s s \rangle \over 128 \pi^2}
-{ \langle g_s^2 GG \rangle \langle g_s \bar s \sigma G s \rangle \over 1536\pi^2}
\\ \nonumber &&
+{11 m_s\langle \bar s s \rangle \langle g_s \bar s \sigma G s \rangle \over32 }\Big )\delta(s)
+ \Big(- { 25 m_s \langle g_s \bar s \sigma G s \rangle^2 \over 384} 
\\ \nonumber &&
+{7 m_s\langle g_s^2 GG \rangle \langle \bar s s \rangle^2 \over 96} - {m_s^2\langle g_s^2 GG \rangle \langle g_s \bar s \sigma G s \rangle \over 256\pi^2} \Big)\delta^\prime(s)\, ,
\\
&& \nonumber \rho^{\rm OPE}_{3/2,1}(s)  
\\ &=& \label{OPE31}
\frac{126}{25} \times \bigg[{5 s^3 \over 36864 \pi^4}-{167 m_s^2 s^2 \over 40960 \pi^4}
\\  \nonumber&&
- \Big( {5 \langle g_s^2 GG \rangle \over 49152 \pi^4}
-{11 m_s \langle \bar s s \rangle \over 1024 \pi^2}\Big )s
\\ \nonumber &&
+ \Big ( - {263 m_s \langle g_s \bar s \sigma G s \rangle \over 73728 \pi^2}
+ { 155 m_s^2\langle g_s^2 GG \rangle  \over 196608 \pi^4 }\Big )
\\ \nonumber &&
-\Big ({89 \langle \bar s s \rangle \langle g_s \bar s \sigma G s \rangle \over 3072}
-{29 m_s\langle g_s^2 GG \rangle \langle \bar s s \rangle \over 147456 \pi^2}
\\ \nonumber &&
-{m_s^2\langle \bar s s \rangle^2\over128 } \Big )\delta(s)
+ \Big( -{361 \langle g_s \bar s \sigma G s \rangle^2 \over 36864} 
\\ \nonumber &&
-{\langle g_s^2 GG \rangle \langle \bar s s \rangle^2 \over 1152} + { m_s\langle g_s^2 GG \rangle \langle g_s \bar s \sigma G s \rangle \over 3072\pi^2}
\\ \nonumber &&
+{7 m_s^2 \langle \bar s s \rangle \langle g_s \bar s \sigma G s \rangle \over 192} \Big)\delta^\prime(s)\bigg]\, ,
\\
&& \nonumber \rho^{\rm OPE}_{3/2,0}(s)  
\\ &=& \label{OPE30}
\frac{126}{25} \times \bigg[{23 m_s s^3 \over 32768 \pi^4}-{\langle \bar s s \rangle s^2 \over 96 \pi^2}
- \Big( {317 m_s \langle g_s^2 GG \rangle \over 589824 \pi^4}
\\  \nonumber&&
-{103 m_s^2 \langle \bar s s \rangle \over 1536 \pi^2}
+{37 \langle g_s \bar s \sigma G s \rangle \over 3072 \pi^2} \Big )s
\\ \nonumber &&
+ \Big (  {441 m_s^2 \langle g_s \bar s \sigma G s \rangle \over 8192 \pi^2}
+ { 25 \langle \bar s s \rangle \langle g_s^2 GG \rangle  \over 24576 \pi^2 }\Big )
\\ \nonumber &&
-\Big ({13 m_s^2\langle g_s^2 GG \rangle \langle \bar s s \rangle \over 4096 \pi^2}
-{13 \langle g_s^2 GG \rangle \langle g_s \bar s \sigma G s \rangle \over 294912\pi^2}
\\ \nonumber &&
+{5 m_s\langle \bar s s \rangle \langle g_s \bar s \sigma G s \rangle \over64 }\Big )\delta(s)
+ \Big( -{5 m_s \langle g_s \bar s \sigma G s \rangle^2 \over 2084} 
\\ \nonumber &&
-{3m_s\langle g_s^2 GG \rangle \langle \bar s s \rangle^2 \over 1024} + { 9m_s^2\langle g_s^2 GG \rangle \langle g_s \bar s \sigma G s \rangle \over 8192\pi^2} \Big)\delta^\prime(s)\bigg]\, .
\end{eqnarray}
Note that we multiply an overall factor to the spectral density of the spin-$3/2$ state . This modification plays a crucial role in the calculation of the mass splitting, as will be discussed in detail later. 

By matching the spectral densities at both the hadron and quark-gluon levels, and applying the Borel transformation, we derive the sum rules as follows:
\begin{eqnarray}
\Pi_{\xi,\mp}(s_0,M_B)&=& 2M_{\xi,\mp}f_{\xi,\mp}^2 e^{-M^{2}_{\xi,\mp}/M_B^2} \label{sumrule}
\\ \nonumber &=& \int^{s_0}_{s_<} (\sqrt{s} \rho^{\rm OPE}_{\xi,1}(s) \pm \rho^{\rm OPE}_{\xi,0}(s))e^{-s/M_B^2}ds .
\end{eqnarray}
and further obtain the formulae for mass and coupling constant:
\begin{eqnarray}
\label{eq:mass} && M^2_{\xi,\mp}(s_0, M_B) 
\\ \nonumber&=& \frac{\int^{s_0}_{s_<} (\sqrt{s} \rho^{\rm OPE}_{\xi,1}(s) \pm \rho^{\rm OPE}_{\xi,0}(s)) s e^{-s/M_B^2} ds}{\int^{s_0}_{s_<} (\sqrt{s} \rho^{\rm OPE}_{\xi,1}(s) \pm \rho^{\rm OPE}_{\xi,0}(s)) e^{-s/M_B^2} ds} ,
\end{eqnarray}
and
\begin{eqnarray}
\label{eq:decay} &&f^2_{\xi,\mp}(s_0, M_B) 
\\\nonumber&=&\frac{\int^{s_0}_{s_<} (\sqrt{s} \rho^{\rm OPE}_{\xi,1}(s) \pm \rho^{\rm OPE}_{\xi,0}(s)) e^{-s/M_B^2} ds \times e^{M_{\xi,\mp}^2/M_B^2}}{2M_{\xi,\mp}} .
\end{eqnarray}
In this way, we calculate the masses and coupling constants of both positive- and negative-parity states as reported in our previous work~\cite{Su:2024lzy}, with the results summarized in the Table~\ref{tab:results}. We find that currents containing the derivative operator couple better to the negative-parity states, and the calculations for states with different parities are performed independently. Therefore, the contributions from the positive-parity states are neglected in the subsequent calculation of the mass splitting. 

Now assuming that the two spin states split by a fine splitting due to spin-orbit dynamics, let us define
\begin{eqnarray}
\Pi_{3/2^-}(q^2) = \Pi_{1/2^-}(q^2) + \Pi_{\delta}(q^2)\, ,
\end{eqnarray}
both at the hadron and quark–gluon levels. At the hadron level, the correlation functions are expressed as:
\begin{eqnarray}
\Pi^{\rm phen}_{3/2^-}(q^2)&\!\!=\!\!&f_{3/2^-}^2 \frac{\slashed{q}+M_{3/2^-}}{M_{3/2^-}^2-q^2 - i \epsilon}
\, ,
\\
\Pi^{\rm phen}_{1/2^-}(q^2)&\!\!=\!\!&f_{1/2^-}^2 \frac{\slashed{q}+M_{1/2^-}}{M_{1/2^-}^2-q^2 - i \epsilon}\,.
\end{eqnarray}
Introducing the mass and coupling differences as $M_{3/2^-}=M_{1/2^-}+\delta M$, $f_{3/2^-}=f_{1/2^-}+\delta f$, we obtain
\begin{eqnarray}
&&f_{3/2^-}^2 \frac{\slashed{q}+M_{3/2^-}}{M_{3/2^-}^2-q^2}
\\ \nonumber &=&(f_{1/2^-}+\delta f)^2 \frac{\slashed{q}+(M_{1/2^-}+\delta M)}{(M_{1/2^-}+\delta M)^2-q^2}
\\ \nonumber
&=&f_{1/2^-}^2
\frac{\slashed{q} + M_{1/2^-}}{M_{1/2^-}^2 - q^2}
\\ \nonumber
&+& f_{1/2^-}^2 \delta M \frac{(M_{1/2^-}^2 - q^2)-2M_{1/2^-} (\slashed{q} + M_{1/2^-})}{(M_{1/2^-}^2 - q^2)^2}
\\ \nonumber
&+&2f_{1/2^-}\delta f
\frac{\slashed{q} + M_{1/2^-}}{M_{1/2^-}^2 - q^2}
+\cdots \,.
\end{eqnarray}
Here we only consider the first order corrections with respect to $\delta M$ and $\delta f$. From the above expression, we can extract the difference in the correlation function $\Pi_{\rm \delta}^{\rm phen}$ as:
\begin{eqnarray}
&&\Pi^{\rm phen}_{\delta}(q^2) 
\\&=& f_{1/2^-}^2 \delta M \frac{-2M_{1/2^-}\slashed{q}-M_{1/2^-}^2 - q^2}{(M_{1/2^-}^2 - q^2)^2}\nonumber
\\ \nonumber
&+&2f_{1/2^-}\delta f\frac{\slashed{q} + M_{1/2^-}}{M_{1/2^-}^2-q^2} \,,
\end{eqnarray}
and reorganize it into two parts with different structures:
\begin{eqnarray}
\Pi^{\rm phen}_{\delta,1}(q^2)\slashed{q}&=&\Big (2f_{1/2^-}\delta f\frac{1}{M_{1/2^-}^2-q^2}
\\ \nonumber
&-&2 f_{1/2^-}^2 M_{1/2^-} \delta M\frac{1}{(M_{1/2^-}^2-q^2)^2}\Big)\slashed{q}\,,
\end{eqnarray}
\begin{eqnarray}
\Pi^{\rm phen}_{\delta,0}(q^2)&=&
-f_{1/2^-}^2 \delta M \frac{M_{1/2^-}^2+q^2}{(M_{1/2^-}^2-q^2)^2}
\\ \nonumber
&+&2f_{1/2^-}M_{1/2^-}\delta f\frac{1}{M_{1/2^-}^2-q^2}\,.
\end{eqnarray}

At the quark-gluon level, using the results of Eqs.~(\ref{OPE1},\ref{OPE0},\ref{OPE31},\ref{OPE30}), $\Pi^{\rm OPE}_{\delta,1(0)}$ can be derived as:
\begin{eqnarray}
\Pi_{\delta,1}^{\rm OPE}({q^2})=\Pi_{3/2^-,1}^{\rm OPE}(q^2)-\Pi_{1/2^-,1}^{\rm OPE} (q^2)  \,, 
\end{eqnarray}
\begin{eqnarray}
\Pi_{\delta,0}^{\rm OPE}({q^2})=\Pi_{3/2^-,0}^{\rm OPE}(q^2)-\Pi_{1/2^-,0}^{\rm OPE} (q^2)   \,.
\end{eqnarray}
Equating the two results at the hadron and quark-gluon levels and performing the Borel transformation, we arrive at:
\begin{eqnarray}
\rm \hat{\mathcal{B}}\Pi^{\rm OPE}_{\delta,1}\label{eq:md1}(s_0,M_B)&=&2f_{1/2^-}\delta f e^{-M_{1/2^-}^2/M_B^2}
\\ \nonumber 
&-&\frac{2 f_{1/2^-}^2 M_{1/2^-} \delta M}{M_B^2}e^{-M_{1/2^-}^2/M_B^2}\,,
\end{eqnarray}
\begin{eqnarray}
\rm \hat{\mathcal{B}}\Pi^{\rm OPE}_{\delta,0}\label{eq:md2}(s_0,M_B)&=&2f_{1/2^-}M_{1/2^-} \delta f e^{-M_{1/2^-}^2/M_B^2}
\\ \nonumber
&+&f_{1/2^-}^2(1-\frac{2M^2_{1/2^-}}{M_B^2}) \delta M e^{-M_{1/2^-}^2/M_B^2}\,.
\end{eqnarray}
By solving Eqs.~(\ref{eq:md1},\ref{eq:md2}) simultaneously, we extract the quantities $\delta M$ and $\delta f$. Note that in the above calculation of the mass splitting, we modify $\Pi_{3/2^-}^{\rm OPE}(q^2)$ by multiplying it with an overall factor such that its perturbative term is identical to that of $\Pi_{1/2^-}^{\rm OPE}(q^2)$ as shown in Eqs.~(\ref{OPE31},\ref{OPE30}). This technique is equivalent to normalizing the currents. It does not change the extracted mass according to Eq.~(\ref{eq:mass}), but affects the coupling constant in Eq.~(\ref{eq:decay}). Therefore, the derived $\delta M$ represents the genuine mass splitting between the two spin-orbit states, whereas the $\delta f$ does not carry the same physical meaning.

To determine the values of $\delta M$ and $\delta f$, we require the parameters specified in Eqs.~(\ref{eq:md1},\ref{eq:md2}), namely the threshold $s_0$, the Borel Mass $M_B^2$, the quantities $f_{1/2}$ and $M_{1/2}$.
In our previous work~\cite{Su:2024lzy}, we carried out a detailed analysis of the masses and coupling constants of the $P$-wave $\Omega$ baryons with quantum numbers $J^{PC}=1/2^-$ and $J^{PC}=3/2^-$, results are summarized in Table.~\ref{tab:results}. For both states, the central values of the input parameters are taken as $s_0=6.0~\rm{GeV}^2$ and $M_B^2=1.65~\rm{GeV}^2$. The extracted mass and coupling constant for $J^{PC}=1/2^-$ state are:
\begin{eqnarray}
M_{1/2^-} &=& 2.07^{+0.07}_{-0.07}{\rm~GeV} \, ,
\label{eq:mass1-}
\\ \nonumber
f_{1/2^-} &=& 0.079^{+0.011}_{-0.011}{\rm~GeV^3} \, .
\label{eq:decay1-}
\end{eqnarray}
Utilizing these inputs, we can straightforwardly determine $\delta M$ and $\delta f$ as:
\begin{eqnarray}
\delta M &=& -0.018^{+0.0336}_{-0.0171}{\rm~GeV} \, ,
\\ 
\delta f &=& 0.00017^{+0.0023}_{-0.0028}{\rm~GeV^3}\, .
\label{eq:mass1-}
\end{eqnarray}
The uncertainties arise from variations in $s_0$, $M_B^2$ as reported in Ref.~\cite{Su:2024lzy} where $5.5
~\rm GeV^2 \leq s_0 \leq 6.5~\rm GeV^2$ and $1.58
~\rm GeV^2 \leq M_B^2 \leq 1.73~\rm GeV^2$, together with the condensates listed below~\cite{ParticleDataGroup:2024cfk,Ovchinnikov:1988gk,Yang:1993bp,Ellis:1996xc,Ioffe:2002be,Jamin:2002ev,Gimenez:2005nt,Narison:2011xe,Narison:2018dcr}:
\begin{eqnarray}
\langle\bar qq \rangle &=& -(0.240 \pm 0.010)^3 \mbox{ GeV}^3 \, ,\label{eq:condensate}
\\ \nonumber \langle\bar ss \rangle &=& (0.8\pm 0.1)\times \langle\bar qq \rangle \, ,
\\ \nonumber \langle g_s\bar s\sigma G s\rangle &=&  (0.8 \pm 0.2)\times\langle\bar ss\rangle \, ,
\\ \nonumber \langle \alpha_s GG\rangle &=& (6.35 \pm 0.35) \times 10^{-2} \mbox{ GeV}^4 \, ,
\\ \nonumber m_s &=& 93 ^{+9}_{-3} \mbox{ MeV} \, .
\end{eqnarray}

For completeness, we have also examined an alternative treatment in which $\delta f=0$ is imposed by redefining the current $J^\prime_\mu=f_{1/2}/f_{3/2}J_\mu$. The resulting mass splitting is found to be similar to that obtained in our main analysis, reinforcing the robustness of our conclusion that the splitting is small.

\begin{table*}[hpt]
\begin{center}
\renewcommand{\arraystretch}{1.25}
\caption{Masses and coupling constants extracted from the currents $J$ in Eq.~(\ref{def:current1}), and $J_{\mu}$ in Eq.~(\ref{def:current2}).}
\begin{tabular}{c|c|c|c|c|c|c|c}
\hline\hline
~\multirow{2}{*}{Current}~& ~~\multirow{2}{*}{state}~~ & ~\multirow{2}{*}{~$s_0^{\rm min}~[{\rm GeV}^2]$~}~ & \multicolumn{2}{c|}{Working Regions} & ~\multirow{2}{*}{Pole~[\%]}~ & ~\multirow{2}{*}{~Mass~[GeV]~}~&~\multirow{2}{*}{~Couple constant~[GeV$^3$]~}~
\\ \cline{4-5}
&&&~~$M_B^2~[{\rm GeV}^2]$~~&~$s_0~[{\rm GeV}^2]$~~&&&
\\ \hline\hline
$J$&$|\Omega;1/2^+\rangle$&9.7&$1.91$-$2.40$&$11.0$&$40$-$55$&$3.05^{+0.21}_{-0.15}$&$0.168^{+0.045}_{-0.040}$
\\
&$|\Omega;1/2^-\rangle$ &5.5&$1.58$-$1.73$&$6.0$&$40$-$47$&$2.07^{+0.07}_{-0.07}$&$0.079^{+0.011}_{-0.011}$
\\
$J_\mu$&$|\Omega;3/2^+\rangle$&10.5&$2.09$-$2.30$&$11.0$&$40$-$46$&$3.13^{+0.27}_{-0.18}$&$0.074^{+0.015}_{-0.009}$
\\
&$|\Omega;3/2^-\rangle$ 
&5.3&$1.54$-$1.76$&$6.0$&$40$-$51$&$2.05^{+0.09}_{-0.10}$&$0.037^{+0.007}_{-0.007}$
\\ \hline\hline
\end{tabular}
\label{tab:results}
\end{center}
\end{table*}

\section{summary and discussion}
\label{sec:summary}

In this work, we have formulated the QCD sum rules to investigate the mass splitting for $P$-wave $\Omega$ baryons based on our previous work~\cite{Su:2024lzy} where $\Omega(2012)$ was identified as the $P$-wave exciation  with $J^P=3/2^-$. There we constructed the currents for the two spin-orbit partner states of $P$-wave $\Omega$ baryons and applied parity-projected QCD sum rules to have extracted their masses as  
$M_{3/2^-} = 2.05^{+0.09}_{-0.10}{\rm~GeV}$ and
$M_{1/2^-} = 2.07^{+0.07}_{-0.07}{\rm~GeV}$ as in Eqs.~(\ref{eq:mass3},\ref{eq:mass1}).
These values have relatively large uncertainties. To improve them, we have formulated a dedicated sum rules directly for the mass splitting. By parametrizing the hadron masses and couplings as $M_{3/2^-}=M_{1/2^-}+\delta M$ and $f_{3/2^-}=f_{1/2^-}+\delta f$, and then expanding the correlation functions in a Taylor series, we obtain expressions for the mass splitting. At the quark-gluon level, the corresponding expressions are derived from the calculated spectral densities. Matching the two representations, we extract the mass splitting as:
\begin{eqnarray}
\delta M &=& -18.0^{+33.6}_{-17.1}{\rm~MeV} \, ,
\label{eq:mass1-}
\end{eqnarray}
Through this procedure, we obtain an improved prediction of small mass splitting between the $P$-wave $\Omega$ baryons. Considering expecting decay patterns of the $J^P=3/2^-$ and $J^P=1/2^-$ states, where $J^P=1/2^-$ state has generally a large width, the resulting two states may be observed experimentally as one peak of $J^P=3/2^-$ state and a broad bump due to the $J^P=1/2^-$ state. The results indicate that the $\Omega(2109)$ is unlikely to be the spin–orbit partner of the $\Omega(2012)$, since their experimental gap of about 100 MeV is significantly larger than our calculated splitting.

\section*{Acknowledgments}
 
This work is partly supported by
the National Natural Science Foundation of China under Grant No.~12505156 and No.~12075019, the startup research fund of Taiyuan University of Technology under Grant No.~RY2500003238, the Jiangsu Provincial Double-Innovation Program under Grant No.~JSSCRC2021488,
and
the Fundamental Research Funds for the Central Universities. 
P.G. is supported by KAKENHI under Contract No. JP22H00122 and JP25H00400. 
A.H. is supported in part by the Grants-in-Aid for Scientific Research [Grant No. 24K07050(C)].

\bibliographystyle{elsarticle-num}
\bibliography{ref}

\end{document}